\documentclass{PoS}
\usepackage[capbesideposition=outside,capbesidesep=quad]{floatrow}
\usepackage{threeparttable}

\title{MAGIC eyes to the extreme: testing the blazar emission models on EHBLs}

\ShortTitle{MAGIC eyes to the extreme}

\author{\speaker{E. Prandini}, C. Arcaro, V. Fallah Ramazani, K. Asano, G. Bonnoli, M. Cerruti, F. D'Ammando, L. Foffano, and F. Tavecchio for the MAGIC Collaboration\footnote{\texttt{https://magic.mpp.mpg.de/}. For collaboration list see PoS(ICRC2019)1177}\\
        INAF Osservatorio Astronomico di Padova \& INFN Sezione di Padova\\
        E-mail: \email{elisa.prandini@inaf.it}}
\abstract{Extreme high-energy peaked BL Lac objects (EHBLs) are blazars whose synchrotron emission peaks at exceptionally high energies, above few keV, in the hard X-ray regime. So far, only a handful of those objects has been detected at very high energy (VHE, E > 100 GeV) gamma rays by Imaging Atmospheric Cherenkov Telescopes. Very remarkably, VHE observations of some of these blazars (like 1ES 0229+200) have provided evidence of a VHE gamma-ray emission extending to several TeV, which is difficult to explain with standard, one-zone synchrotron self-Compton models  usually applied to BL Lac objects. 
The MAGIC collaboration coordinated a multi-year, multi-wavelength observational campaign on ten targets. The MAGIC telescopes detected VHE gamma rays from four EHBLs. In this paper we focus on the source 1ES~1426+426 and its X-ray and VHE gamma-ray properties. The results of different models (synchrotron self-Compton, spine-layer, hadronic) reproducing the broadband spectral energy distribution are also presented.}

\FullConference{36th International Cosmic Ray Conference -ICRC2019-\\
		July 24th - August 1st, 2019\\
		Madison, WI, U.S.A.}

\begin{document}

\section{Introduction}
Blazars are the most numerous  gamma-ray sources in the extragalactic sky. They are jetted active galaxies whose jet is closely aligned to the line of sight \cite{2016AARv..24...13P}.
Their electromagnetic spectral energy distribution (SED) is  dominated by non-thermal emission and shows two humps: one at low energies  that peaks in the optical-to-X-ray band, attributed to synchrotron radiation by accelerated electrons. The second one at higher energies  whose emission origin, if leptonic and/or hadronic, is still largely debated. 

Among blazars, extreme high-energy peaked BL Lac objects (EHBLs)  form an emerging sub-class featuring extremely energetic emission whose low-energy peak lies above 10$^{17}$\,Hz, in the soft-to-hard X-ray band \cite{2001AA...371..512C,2018MNRAS.477.4257C}.
In some objects also the high-energy peak is located at unusual energies, above 1\,TeV. This is the case of {\it hard-TeV blazars}, that show a sub-TeV emission in the 100\,GeV to 1\,TeV range particularly hard when compared to other blazars, with spectral indices $\Gamma$ harder  than 2 when fitted with a power-law model (dN/dE = F$_0$ (E/E$_0$)$^{-\Gamma}$) indicating indeed an extreme location of  this  peak. These objects are also relatively faint when compared to other gamma-ray blazars. Prototype of this class of sources is 1ES~0229+200 \cite{2001AA...371..512C,aharonian07,tavecchio09}.

Given the extreme conditions characterizing the jet of EHBLs and producing such peculiar emission, these objects represent the ideal laboratory for testing theories of particle acceleration and emission mechanisms \cite{kaufmann11}. Also, in the hard-TeV  cases, the reservoir of $>$TeV photons represents an opportunity of testing with high accuracy models of gamma-ray propagation, in particular the gamma-gamma opacity due to the extragalactic background light (EBL, \cite{aharonian07}). 
Finally, they can be used to provide limits to the intergalactic magnetic field strengths \cite{murase12}.

EHBLs represent an indisputable probe for astrophysical, cosmological, and fundamental physics studies \cite{bonnoli15}. These studies are however limited due to the scarce number of hard-TeV EHBLs known to date. With the aim of increasing their number, the MAGIC Collaboration started a multi-year observational campaign of promising targets. 
Here we present the list of targets observed with MAGIC and  then focus on the detailed properties of one of the detected sources, the previously-known TeV emitter 1ES ~1426$+$428.

\begin{table}
\centering
\caption{Sample of EHBLs observed with MAGIC. $\dagger$ Synchrotron peak frequency reported by \cite{Chang17}.}\label{tab:sources}
\begin{tabular}{lccccc} 
\hline
{Source} & RA (J2000)   & DEC (J2000)  & {$z$} & $\mathrm{log}$($\nu_{\rm{peak}}$)$^\dagger$  \\
                         & [$^{\circ}$] & [$^{\circ}$] &       &   [Hz]           \\
\hline
TXS~0210+515     & 33.57   &  51.75   & 0.049    & 17.3    \\
TXS~0637-128     & 100.03  & -12.89   & unknown  & 17.4      \\
BZB~J0809+3455   & 122.41  &  34.93    & 0.082   & 16.6   \\
RBS~0723         & 131.80  &  11.56   & 0.198   & 17.8    \\
1ES~0927+500     & 142.66  &  49.84   & 0.187   & 17.5      \\
RBS~0921         & 164.03  &  2.87    & 0.236   & 17.9      \\ 
1ES~1426+428     & 217.14  &  42.70  & 0.129     & 18.1      \\ 
1ES~2037+521     & 309.85  &  52.33  & 0.053    & N.A.      \\
RGB~J2042+244    & 310.53  &  24.45   & 0.104  & 17.5     \\
RGB~J2313+147    & 348.49  &  14.74   & 0.163  & 17.7      \\
\hline
1ES~0229+200     & 38.20   &  20.29  & 0.140   & 18.5         \\
\hline
\end{tabular}
\end{table}

\section{Very high energy gamma-ray observations}
MAGIC \cite{magicperf_1:2015} is a system of two telescopes detecting gamma rays above 60\,GeV sky from La Palma, Canary Islands, at an altitude of $\sim$2200\,m.
With the aim of characterizing the TeV emission, ten promising targets listed in Table~\ref{tab:sources} were selected for MAGIC observations according to their SED characteristics, in particular the X-ray and/or gamma-ray properties. 
In addition, the source 1ES~0229$+$200 considered as the prototype of hard-TeV EHBLs was also deeply observed with MAGIC. 

\begin{figure}
    \centering
    \includegraphics[width=0.50\textwidth]{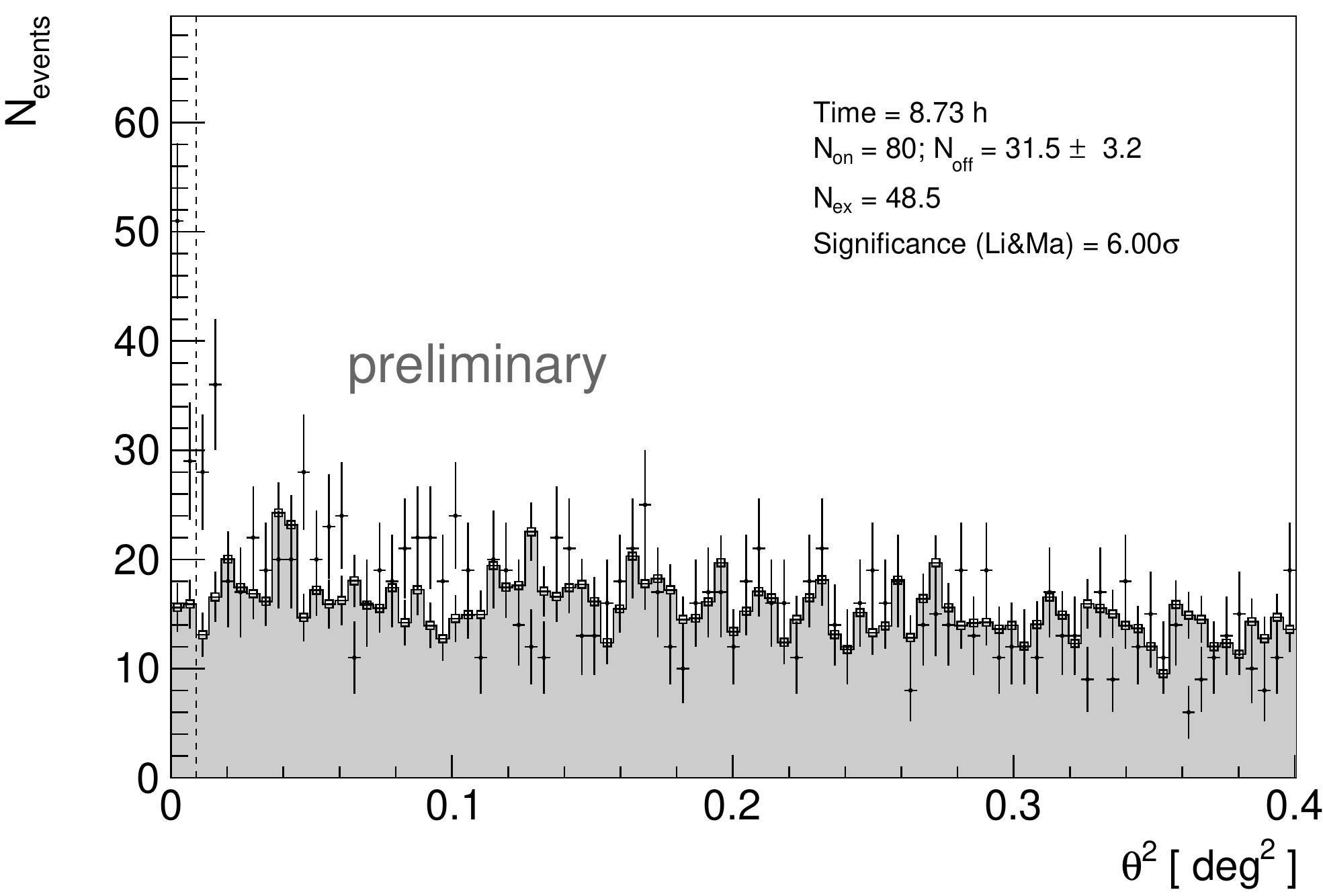}
    \includegraphics[width=0.49\textwidth]{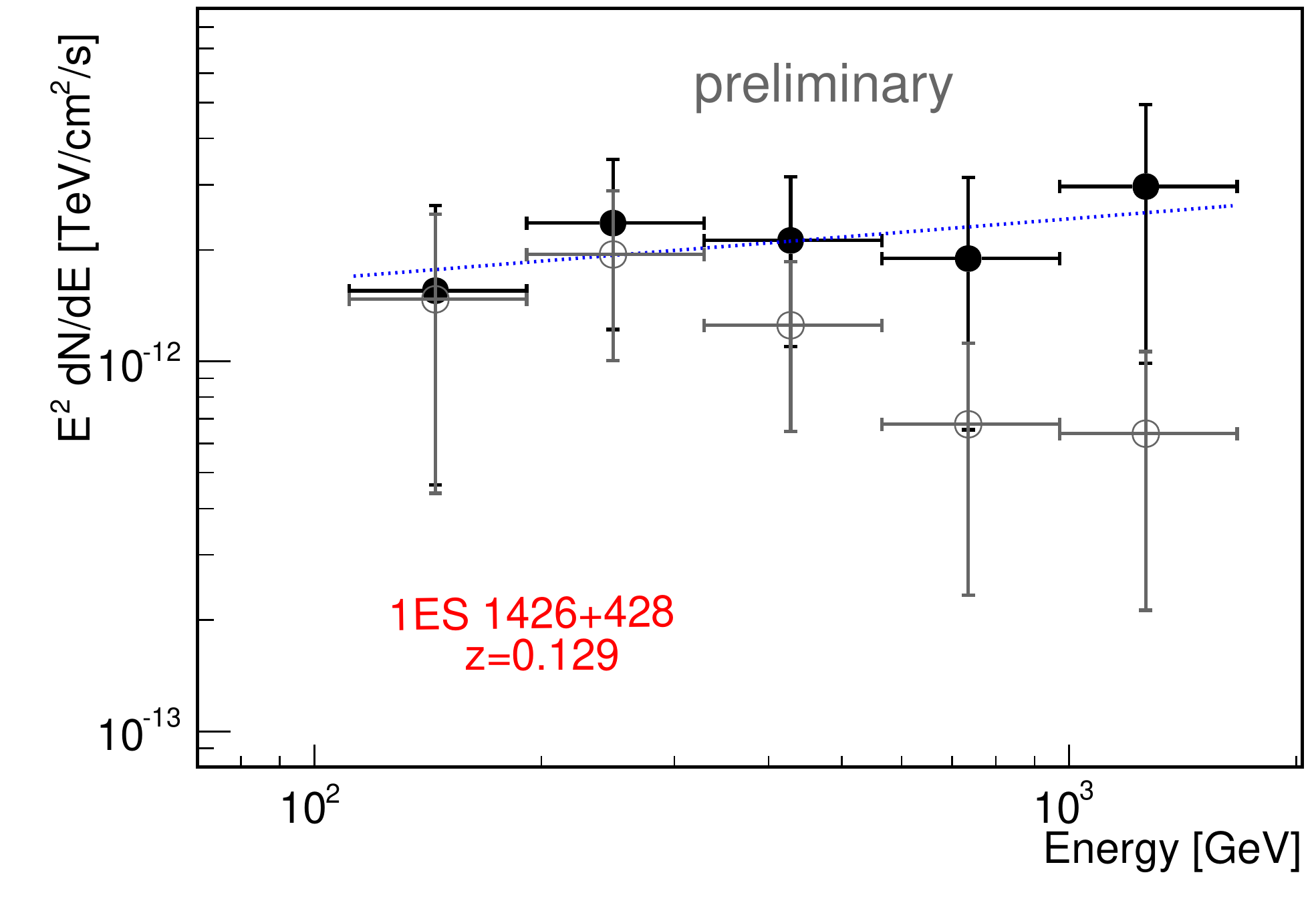}
    \caption{Left: MAGIC detection plot of 1ES~1426+428, for the 2012 observing campaign. Right:  MAGIC observed (open markers) and intrinsic (filled markers) spectrum assuming the EBL model \cite{franceschini08}.}
    \label{fig:1426_detection}
\end{figure}

The overall observations span several years (from 2010 to 2017\footnote{Part of the results of 2018 MAGIC observations of extreme blazars are reported in \cite{foffano_icrc}.}) and the observing conditions were quite variable, with part of the sample collected during partial moon-light and medium-to-large zenith angles (up to 55$^{\circ}$). 
The MAGIC data were analyzed with standard procedures \cite{zanin13}. A significant signal was found from four sources, namely TXS~0210+515, RBS~0723, 1ES~1426+428, and 1ES~2037+521. In addition, a hint of signal was detected from RGB~J2042+244 at a confidence level of $>$3\,$\sigma$ significance.
Among these sources, only 1ES~1426+428 was already known as TeV emitter \cite{2002A&A...384L..23A, 2017ApJ...835..288A}. In Table~\ref{tab:1426_obs} we list the details of 1ES~1426$+$428 observations and the results of the data analysis for the three different MAGIC observation campaigns, in 2010, 2012, and 2013 respectively. 
A signal with 6.0\,$\sigma$ significance was found from the analysis of $\sim$9 hours of 2012 data. The detection plot is reported in Fig.~\ref{fig:1426_detection}. The data from the other two years resulted in non-detection. 
The corresponding 95 per cent confidence upper limits, listed in Table~\ref{tab:1426_obs}, are however compatible with the hypothesis of a constant flux when compared with the detection and when both the systematic ($<15\,\%$) and statistical errors are taken into account.

\begin{table}
\centering
\caption{\label{tab:1426_obs} Results of the signal search and integral flux analysis of the MAGIC data for 1ES~1426+428. Only the statistical uncertainty was considered for the estimation of the 95 per cent confidence upper limits.
}
\begin{tabular}{lccccc}
\hline
 Year  & Exposure Time  & Significance &  Flux$_{E\geq 200\, \mathrm{GeV}}$ \\
             & [h] & $\sigma$            & $\times 10^{-12}$\,[cm$^{-2}$s$^{-1}$] \\
\hline
 2010          & 6.5   & 2.1     & $<$ 9.3  \\
 2012          & 8.7    & 6.0  & 6.1 $\pm$ 1.1  \\  
 2013          & 5.9    & 1.8   & $<$ 5.1 \\ 
\hline
\end{tabular}
\end{table}

The differential energy spectrum resulting from MAGIC 2012 observations is displayed in Fig.~\ref{fig:1426_detection}, right plot open markers, in $E^2dN/dE$ representation. Once corrected for EBL absorption by adopting the model proposed by Franceschini et al. \cite{franceschini08}, the intrinsic spectrum obtained is compatible with a power-law function with spectral index 1.8 $\pm$ 0.5, indicating a hard-TeV 1ES~0229+200-like nature for this source.

\section{X-ray observations}
X-ray observations are crucial to characterize EHBLs, which feature the peak of the synchrotron emission in this band.  Remarkably, the sensitivity of current X-ray telescopes allows  a precise determination of the flux in every single pointing, conversely to what we face in the gamma-ray band. The X-ray light curve, therefore, represents the most precise tool at hands to infer the variability of EHBLs.

\begin{figure}
    \centering
    \includegraphics[width=1.\textwidth]{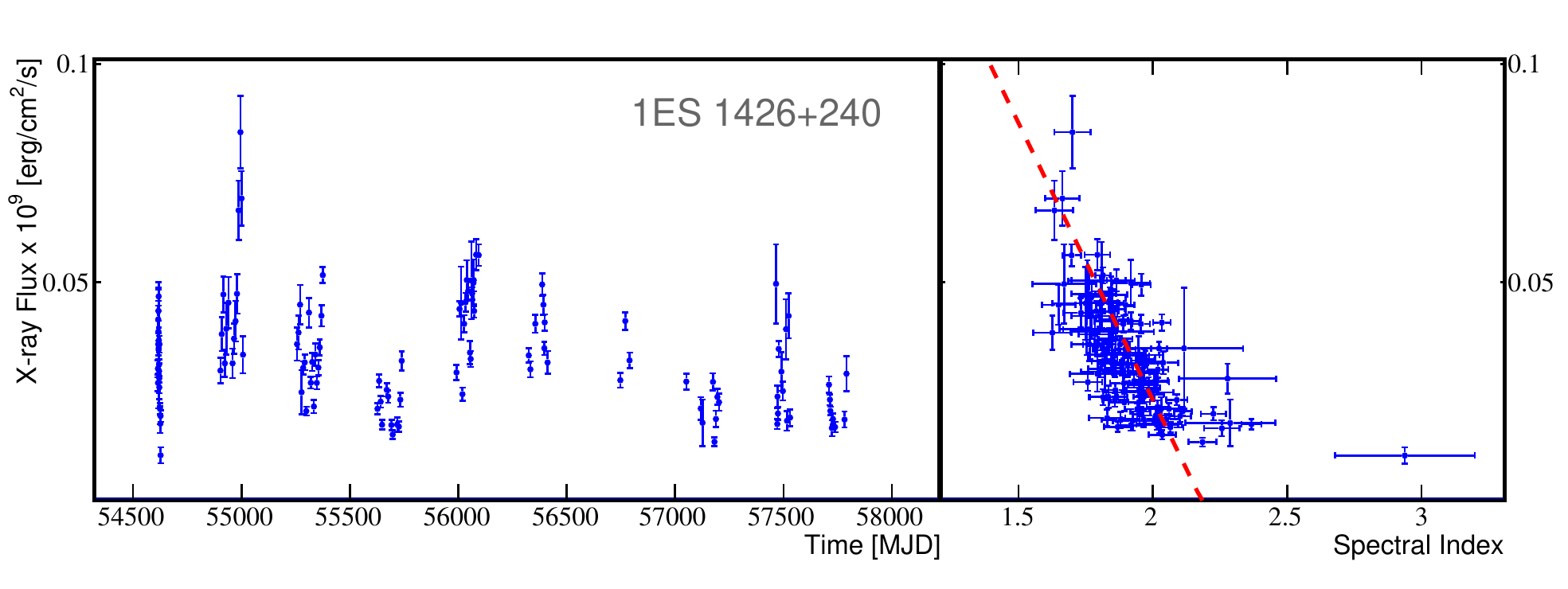}
    \caption{ 1ES 1426+428 X-ray light curve (left panel) and spectral index vs flux scatter plot (right panel).}
    \label{fig:Xray_results}
\end{figure}

As for the other sources of the study  listed in  Table~\ref{tab:sources}, MAGIC observations were complemented with X-ray data collected with the XRT instrument on-board the {\it Neil Gehrels Swift Observatory}. The long-term curve obtained analyzing all the available data with the standard analysis is displayed in Fig.~\ref{fig:Xray_results}, left panel. 
The flux amplitude is variable within a factor of $\sim$8  over $\sim$9 years of monitoring. In 2012, during MAGIC observation window, the flux amplitude is basically constant.
The source shows a harder-when-brighter behavior, as depicted in the right panel of Fig~\ref{fig:Xray_results}. This trend is quite common in blazars and is usually interpreted as the emergence of a higher-energy electron population emitting the synchrotron radiation. 1ES~1426+428 is the most variable source in the X-ray band of our sample.

\section{Modelling}

\begin{figure}
    \centering
    \includegraphics[width=1.\textwidth]{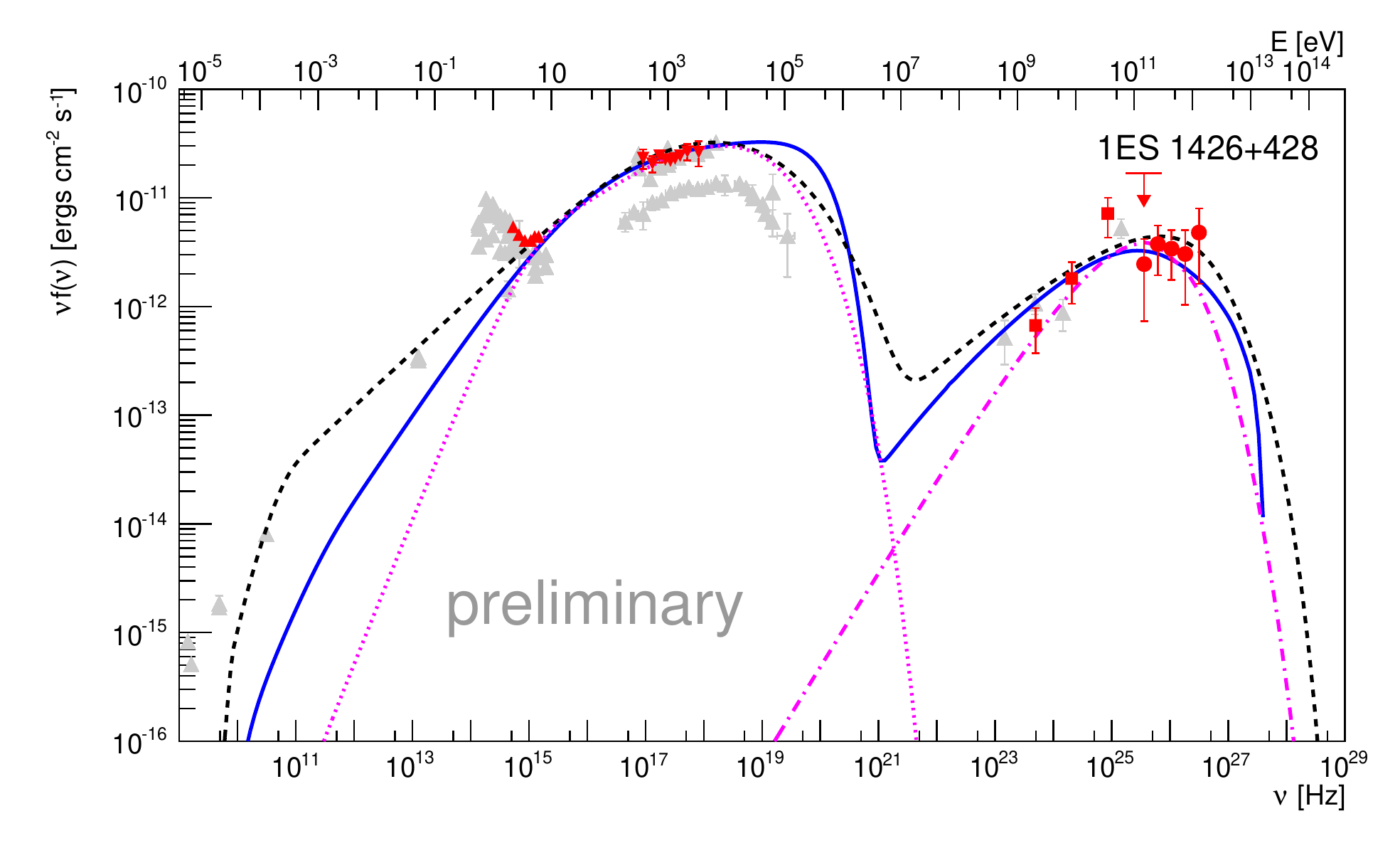}
    \caption{Multi-wavelength SED of 1ES~1426+428 from radio to VHE gamma-ray frequencies  during the 2012 MAGIC campaign. Quasi-simultaneous data are represented with red markers while gray symbols are archival data. The lines represent the SSC model (continuous blue line), spine-layer model (black dashed line), and lepto-hadronic model (magenta dotted line for the leptonic component, dashed-dotted line for the hadronic component). See text for details.}
    \label{fig:SED}
\end{figure}

Figure~\ref{fig:SED} illustrates the  broadband SED built with quasi-simultaneous data collected in 2012, red markers, superimposed to archival data from the ASI Science Data Center (ASDC), gray markers.
Quasi-simultanoeus observations include {\it Swift}-UVOT data from MJD~56064 (17 May 2012) covering the optical-UV band, which show a clear evidence of host-galaxy emission at the lower frequencies. 
The {\it Swift}-XRT spectrum from the same day is well fitted by a power-law function with index $1.84 \pm 0.02$. 
The {\it Fermi}-LAT spectrum in the GeV band obtained from a dedicated data analysis centered on the interval MJD~55927--56292 (1 January 2012 -- 31 December 2012) is also well fitted by a power-law function with index $1.4 \pm 0.2$. 
The large time span considered for {\it Fermi}-LAT data is motivated by the faintness of the signal  at these frequencies.
Finally, MAGIC EBL-corrected data presented in the previous section are displayed.
The location of both SED peaks remains unconstrained, as no clear curvature is observed  neither  in the soft X-ray nor in the VHE band, suggesting an extreme location of both synchrotron and high-energy peak.

\begin{table}
\floatbox[\capbeside]{table}
	{	\caption{Model parameters and obtained physical values for the SSC conical-jet scenario applied to the SED of 1ES~1426+428. 
    From top to bottom: bulk Lorentz factor of the jet; magnetic field; electron luminosity; maximum electron Lorentz factor; spectral index of the electron energy distribution; synchrotron and inverse Compton peak energy resulting from the model; Compton-dominance parameter (CD); ratio between the magnetic and electron energy density evaluated at the radius where the electron injection shuts down.
        }	\label{param_tab}}
{		\begin{tabular}{lccccccccccc}
			\hline
			\hline
			 $\Gamma$ & &  $20$ \\
			 $B_0$ & [G] & 0.20 & \\
			 $L_{\rm e}$ & [$\mbox{erg}~\mbox{s}^{-1}$] & $1.3 \times 10^{44}$ \\
			 $\gamma_{\rm max}$ & &  $2 \times 10^{6}$ \\
			 $n_1$ & &  2.0 \\
			$\mathrm{log}$($\nu_{\rm{syn,pk}}$) & [Hz] & 18.2 \\
			$\mathrm{log}$($\nu_{\rm{IC,pk}}$) & [Hz] &  25.8 \\
			CD & & 0.14 \\
			$U_B$/$U_{\rm e}$ & & $2.6 \times 10^{-2}$ \\
         \hline
         \hline
		\end{tabular}}
\end{table}

\begin{table}
\floatbox[\capbeside]{table}
{\caption{Model  parameters  and  obtained  physical  values  for  the  spine-layer scenario applied to  the SED of 1ES~1426+428.
From top to bottom: minimum, break and maximum electron Lorentz factor. Spectral index of the electron energy distribution below and above $\gamma _b$; magnetic field; normalization of the electron distribution; radius of the emission zone; Doppler factor; magnetic energy density; relativistic electron energy density; kinetic luminosity of the jet.}\label{tab:spine}}
{\begin{tabular}{lcc}
\hline
\hline
$\gamma _{\rm min}$ &  & $100 $ \\
$\gamma _{\rm b}$&  & $ 3\times 10^{4} $ \\
$\gamma _{\rm max}$&  &$ 2\times 10^{6} $\\
$n_1$&  & $ 1.4 $ \\
$n_2$ &  & $ 2.9 $ \\
$B$ & [G] &  $ 0.34 $ \\ 
$K$ & cm$^{-3}$ & $ 3.5$ \\
$R$ & $10^{15}$ cm &$ 7.1 $ \\
$\Gamma_s $=$\delta $ &  & 20  \\
$U_B$ & erg cm$^{-3}$ & $4.6\times 10^{-3}$\\ 
$U_{\rm e}$ & erg cm$^{-3}$ & $4.3\times 10^{-3}$ \\
$L_{\rm j}$ & $10^{42}$ erg s$^{-1}$ & 20.5 \\
\hline
\hline
\end{tabular}}
\end{table}

\begin{table}
    \floatbox[\capbeside]{table}
{\caption{Model  parameters  and  obtained  physical  values  for the hadronic scenario  applied to the SED of 1ES~1426+428. The quantities flagged with a star ($^\star$) are derived quantities, and not model parameters.}    \label{tab:hadronic-model}}
 {\begin{tabular}{lcc}
   		\hline
   		\hline
 		$\delta$ & &30 \\
        $R$ &[10$^{16}$ cm] &  $0.01-13.8$\\
     $^\star \tau_\textnormal{obs}$ &[hours] & $0.03-48.0$ \\
 		\hline
 		 $B$ &[G] & $2.0-344$   \\
 		$^\star u_B$ &[erg cm$^{-3}$] & $0.17-4710$ \\
 		\hline
 		$\gamma_{e,\textnormal{min}} $& & $200$ \\
 		$\gamma_{e,\textnormal{b}} $& & $=\gamma_{e,\textnormal{min}}$ \\
 		$\gamma_{e,\textnormal{max}}$&$ [10^4]$  & $1.2-15.9$ \\
 		$n_{e,1}$=$n_{p,1}$& & $1.25$  \\
 		$n_{e,2}$=$n_{p,2}$& & $2.25$  \\
 		$K_e$ &[10$^{-3}\, $cm$^{-3}$] & $0.09-1.2\times10^5$   \\
 		$^\star u_e$ &[10$^{-7}\,$erg$\,$cm$^{-3}$] & $0.12-1.2\times10^5$  \\
 		\hline
 		$\gamma_{p,\textnormal{min}}$&& 1 \\
 		$\gamma_{p,\textnormal{break}}$ & $ [10^9]$ &  $=\gamma_{p,\textnormal{max}}$ \\
 		$\gamma_{p,\textnormal{max}}$&$ [10^9]$&  $1.6-21.0$ \\
         $\eta$ & [10$^{-5}$]& $0.07-1.7$\\
 		$^\star u_p$ & [10$^{-4}\,$erg cm$^{-3}$] & $0.06-2.7\times10^5$  \\
 		\hline
    	$^\star u_p$/$u_B $ &[10$^{-5}$] &  $2.8-1070$\\
 		$^\star L$& [10$^{46}$ erg s$^{-1}$] &  $0.11-18.2$  \\
 		\hline
 		\hline
 		\end{tabular}}
 		\end{table}
 		
For the interpretation of the  broadband emission, we consider three different models: two leptonic models, namely a synchrotron self-Compton model and a spine-layer model, and a proton-synchrotron lepto-hadronic model. 		

 All these models provide a good description of the SED. The model parameters are however substantially different resulting in three different scenarios for 1ES~1426+428:
\begin{description}
\item[SSC model -]  the SSC conical-jet scenario described in \cite{2014ApJ...780...64A} and applied to our sample interprets the synchrotron peak  as due to the cooling break of the electron parent population. In other EHBLs, instead, a break in the electron distribution is assumed to fit the data.  The model results in a synchrotron peak at 10$^{18.2}\,Hz$ and an inverse Compton peak at 10$^{25.2}$\,Hz. 
Main result is a low magnetization parameter, even if it is larger than what usually found in other hard-TeV EHBLs. The Compton-dominance parameter value, i.e. the ratio of $\nu L_\nu$ at the IC peak to that at the synchrotron peak, is 0.14, similar to other objects of this class.    

\item[Spine-layer model -] the spine-layer model described in \cite{2005A&A...432..401G,tavecchio16} assumes the existence of two regions in the jet: a faster inner core (the spine, with Lorentz factor $\Gamma_s$), surrounded by a slower sheath of material (the layer, with Lorentz factor $\Gamma_L$). In this work, the Lorentz factors of the spine and the layer are fixed to $\Gamma=20$ and $\Gamma_L=3$. We also assume $\delta=\Gamma$. The other parameters (in particular the luminosity of the layer emission) were varied so that the spine is close to equipartition.  In Table~\ref{tab:spine} we list the parameters used for the spine. As expected, the  magnetic field adopted in this model is higher than the one assumed in the SSC model.

\item[Lepto-hadronic model -] we assume that proton synchrotron radiation as detailed in \cite{cerruti15} is responsible for the gamma-ray component of the blazar SED. The number of free parameters of hadronic blazar models is much higher than the one of leptonic models, so we make the following physically motivated assumptions to reduce the parameter space: (1) the Doppler factor of the emitting region $\delta$ is fixed to 30, a quite common value in BL Lacs; (2) $R\leq 1.6\times10^{17}\ (1+z)^{-1} {\rm cm}$ is assumed, which is in agreement with  the absence of a fast (day-scale or less) variability detection; (3) for protons, the efficiency of the acceleration mechanism is fixed to $0.1$ and the cooling time-scales considered are the adiabatic and the synchrotron ones; (4) for leptons the main cooling mechanism is assumed to be synchrotron radiation. The results of our fit are listed in Table~\ref{tab:hadronic-model}.  In particular, the luminosity of the emitting region has been calculated as $L=2 \pi R^2c\Gamma_{bulk}^2(u_B+u_e+u_p)$, where $\Gamma_{bulk}=\delta/2$, and $u_B$, $u_e$, and $u_p$  are the energy densities of the magnetic field, the electrons, and the protons, respectively. With respect to leptonic models, here we obtain a strong magnetic field and a consequently very high magnetization. As an outcome of the model we also consider the neutrino flux, which lies well below the detection capabilities of current neutrino telescopes \cite{icecube}.
\end{description}


\section{Conclusions}
In this paper, we presented the results of the VHE gamma-ray and X-ray analysis for the EHBL 1ES~1426+428. This source is part of a large program carried out by the MAGIC Collaboration aimed at studying and characterizing the emission of EHBLs at VHE. Both the VHE and the X-ray spectra are hard, with only modest variations in the X-ray band over nine years of monitoring.

The broadband SED built with quasi-simultaneous data leave the synchrotron and high-energy peaks of the SED still largely unconstrained. Both the  simple SSC conical-jet and structured-jet leptonic model, as well as the proton-synchrotron hadronic model, propose an extremely energetic synchrotron peak for this source exceeding 10$^{18}$\,Hz, and a not so extreme high-energy peak, in the sub-TeV energy range. Further observations of the source in these energetic bands are crucial to better constrain the SED shape.

Main outcome of the considered models  is that while the data can be well-reproduced in all cases, only the spine-layer scenario is able to propose a solution with physical conditions close to equipartition. The other two models instead provide two opposite solutions, with a highly magnetized condition in the hadronic case and a very weak magnetic field in the SSC one. This agrees with the modelling results of the other EHBLs studied by the MAGIC Collaboration.

\section*{Acknowledgements}
%
%
This project has received funding from the European Union's Horizon2020 research and innovation programme under the Marie Sklodowska--Curie grant agreement no 664931.
Part of this work is based on archival data, software, or online services provided by the Space Science Data Center - ASI. This research has made use of data and/or software provided by the High Energy Astrophysics Science Archive Research Center (HEASARC), which is a service of the Astrophysics Science Division at NASA/GSFC and the High Energy Astrophysics Division of the Smithsonian Astrophysical Observatory.
We would like to thank the Instituto de Astrof\'{\i}sica de Canarias for the excellent working conditions at the Observatorio del Roque de los Muchachos in La Palma. The financial support of the German BMBF and MPG, the Italian INFN and INAF, the Swiss National Fund SNF, the ERDF under the Spanish MINECO (FPA2015-69818-P, FPA2012-36668, FPA2015-68378-P, FPA2015-69210-C6-2-R, FPA2015-69210-C6-4-R, FPA2015-69210-C6-6-R, AYA2015-71042-P, AYA2016-76012-C3-1-P, ESP2015-71662-C2-2-P, FPA2017-90566-REDC), the Indian Department of Atomic Energy, the Japanese JSPS and MEXT, the Bulgarian Ministry of Education and Science, National RI Roadmap Project DO1-153/28.08.2018 and the Academy of Finland grant nr. 320045 is gratefully acknowledged. This work was also supported by the Spanish Centro de Excelencia ``Severo Ochoa'' SEV-2016-0588 and SEV-2015-0548, and Unidad de Excelencia ``Mar\'{\i}a de Maeztu'' MDM-2014-0369, by the Croatian Science Foundation (HrZZ) Project IP-2016-06-9782 and the University of Rijeka Project 13.12.1.3.02, by the DFG Collaborative Research Centers SFB823/C4 and SFB876/C3, the Polish National Research Centre grant UMO-2016/22/M/ST9/00382 and by the Brazilian MCTIC, CNPq and FAPERJ.


\end{document}